\documentclass[journal,onecolumn,11pt]{IEEEtran}
\usepackage{hanw_preamble_trans}

\newcommand{\Prb}{\mathbbmss{P}}

\begin{document}
\title{On Joint Communication and Channel Discrimination}
\author{Han Wu and Hamdi Joudeh \footnotetext{The authors are with the Department of Electrical Engineering, Eindhoven University of Technology, 5600 MB Eindhoven, The Netherlands (e-mail: h.wu1@tue.nl; h.joudeh@tue.nl).}}
\markboth{Draft}
{Shell \MakeLowercase{\textit{et al.}}: Bare Demo of IEEEtran.cls for IEEE Journals}
\maketitle
\begin{abstract}
We consider a basic communication and sensing setup comprising a transmitter, a receiver and a sensor. The transmitter sends an encoded sequence to the receiver through a discrete memoryless channel, and the receiver is interested in decoding the sequence. 
On the other hand, the sensor picks up a noisy version of the transmitted sequence through one of two possible discrete memoryless channels. 
The sensor knows the transmitted sequence and wishes to discriminate between the two possible channels, i.e. to identify the channel that has generated the output given the input. 
We study the trade-off between communication and sensing in the asymptotic regime, captured in terms of the coding rate to the receiver against the discrimination error exponent at the sensor.
We characterize the optimal rate-exponent trade-off for general discrete memoryless channels with an input cost constraint.
\end{abstract}
\section{Introduction}
\label{sec:introduction}
\begin{figure}[h]
\centering
\captionsetup{justification=centering}
\includegraphics[width=0.7\textwidth]{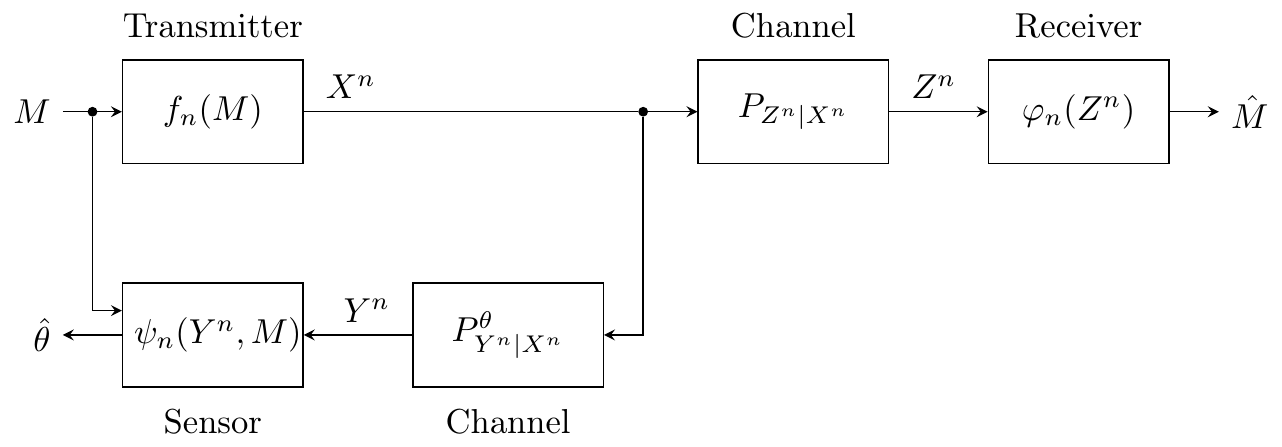}
\caption{An illustration of the considered setting. A precise definition of all blocks is given in Section \ref{sec:problem_setting}.}
\label{fig:block_diagram}
\end{figure}

We consider a setting comprising a transmitter, a receiver and a sensor. 
The transmitter has a random message $M$ which it encodes into a sequence $X^n \triangleq X_1,X_2,\ldots,X_n$ of length $n$, drawn from an alphabet $\mathcal{X}^n$. This sequence serves as input to a pair of channels $P_{Z^n|X^n}: \mathcal{X}^n \to \mathcal{Z}^n$ and $P_{Y^n|X^n}^{\theta} : \mathcal{X}^n \to \mathcal{Y}^n$, where $\mathcal{Z}^n$ and $\mathcal{Y}^n$ are the corresponding output alphabets.
The receiver observes $Z^n \triangleq Z_1,Z_2,\ldots,Z_n$ through $P_{Z^n|X^n}$ and wishes to retrieve the message $M$ from $Z^n$.
The sensor, on the other hand, observes $Y^n \triangleq Y_1,Y_2,\ldots,Y_n$ through $P_{Y^n|X^n}^{\theta}$, which depends on a fixed yet unknown parameter $\theta$ taking values in $\Theta$. 
The sensor has $M$ as side information, and wishes to estimate the channel parameter $\theta$ from  $(Y^n,M)$. 
An illustration of this setting is shown in Fig. \ref{fig:block_diagram}. 

The above setting is a basic model for joint communication and sensing systems. Such systems may be used in the context of automotive communication and radar sensing,  where a vehicle sends a message-bearing signal to another vehicle  and then uses the backscattered echo of the transmitted  signal to identify and track road obstacles, see, e.g. \cite{Sturm2011,Ma2020}. 
Our aim is to shed some light on the fundamental performance limits of such systems.
As a step in this direction, in this paper we focus on discrete memoryless settings: the input and output alphabets are finite, and the noisy channels are stationary and memoryless.
We also limit our attention to a basic sensing task where the parameter $\theta$ is drawn from $\Theta = \{0,1\}$.
That is, with knowledge of $M$ (and hence $X^n$) and upon observing $Y^n$, the sensor wishes to distinguish between the two channels $P_{Y^n|X^n}^{0}$ and $P_{Y^n|X^n}^{1}$.
\subsection{Related work}
An information theoretic formulation for a joint communication and sensing was proposed in 
\cite{Kobayashi2018}, where the authors considered a setting in which a transmitter sends a codeword
to a receiver over a state-dependent memoryless channel, and then estimates 
the channel state sequence from generalized feedback.
The trade-off between message communication and state estimation is characterized in terms of a capacity-distortion function.
The formulation and results of \cite{Kobayashi2018} have been extended to multi-user settings in \cite{Kobayashi2019,Ahmadipour2021b}.
The problems considered in \cite{Kobayashi2018,Kobayashi2019,Ahmadipour2021b} and the one we consider here are similarly motivated yet they differ in their underlying models, which lead to distinct solution approaches.
The channel state process in \cite{Kobayashi2018,Kobayashi2019,Ahmadipour2021b} is i.i.d. while in our formulation the channel parameter $\theta$ remains fixed.
Our model is hence better suited for scenarios where parameters of interest for sensing 
change at a much slower time scale compared to channel symbol periods, e.g. as in radar applications \cite{Sturm2011}.  
Another difference is the availability of generalized feedback in \cite{Kobayashi2018,Kobayashi2019,Ahmadipour2021b}, which is not incorporated in our model.

The basic sensing task that we consider, with a binary parameter $\theta$, is a simple binary hypothesis testing problem. This is a canonical problem in both statistics and information theory, and notable works that characterize the asymptotic performance limits include those by Chernoff \cite{chernoff1952}, Hoeffding \cite{hoeffding1965}, Csisz{\'a}r-Longo \cite{csiszar1971}, and Blahut \cite{Blahut1974}. 
The specific version of the problem that we consider here, where the sensor knows the input $X^n$ and wishes to distinguish between two channels $P_{Y^n|X^n}^{0}$ and $P_{Y^n|X^n}^{1}$ from an observation $Y^n$, is also known as channel discrimination, see, e.g. \cite{Hayashi2009}. This problem has been considered in a number of works under various assumptions, including fixed-length transmission in Blahut \cite{Blahut1974}, fixed-length transmission with feedback (i.e. adaptive) in  Hayashi \cite{Hayashi2009}, and variable-length transmission with feedback in  Polyanskiy-Verd{\'u} \cite{polyanskiy2011_ITA}.
In this paper, we consider the non-adaptive (i.e. no feedback) setting with fixed-length transmission. 
Moreover, in addition to facilitating channel discrimination at the sensor, the input sequence $X^n$ in our setting must also carry a message to the receiver, which distinguishes our problem from the ones previously considered in the literature.
\subsection{Contribution}
We consider the setting illustrated in Fig. \ref{fig:block_diagram} with discrete memoryless channels, a binary parameter $\theta$, and an average input cost constraint; and we study the trade-off between reliable message communication and efficient channel discrimination in the asymptotic regime (i.e.  $n \to \infty$).
This trade-off is captured in terms of the message communication rate against the channel discrimination error exponent. 

We first adopt a maximum (i.e. worst-case) error criterion for channel discrimination, where the goal is to minimize the worst of the two types of error, and we characterize the optimal rate-exponent trade-off region in this case (Theorem \ref{thm:disc_cost}, Section \ref{sec:comm_disc_trade_off}). 
The achievability part of our result is obtained by adapting standard error bounding techniques and a channel coding argument based on strong typicality. For the converse part, the main ingredients are the error lower bound in Shannon-Gallager-Berlekamp \cite{shannonLowerBoundsError1967} and a type-counting argument from Csisz{\'a}r-K{\"o}rner \cite{csiszarInformationTheoryCoding2011}.
Then we adopt a Neyman-Pearson channel discrimination error criterion, where the goal is to minimize one type of error while keeping the other type below a set threshold, and we derive the optimal trade-off in this case as well (Theorem \ref{thm:disc_stein}, Section \ref{sec:Neyman_Pearson}). This case is relevant in many practical applications, e.g. in obstacle detection to avoid road collisions, a missed detection is much worse than a false alarm.
Finally, it is worthwhile mentioning that a special case of the above problem, with binary channels and an on-off channel parameter, was recently considered in \cite{Joudeh2021} under a maximum channel discrimination error criterion. The results in the present paper generalize the one \cite{Joudeh2021} to arbitrary discrete memoryless channels with input cost, and to the Neyman-Pearson error criterion.
\subsection{Notation}
Upper-case letters, e.g. \(X,Y,Z,M\), often denote random variable and the corresponding lower-case letters, e.g. \(x,y,z,m\), denote their realizations. Calligraphic letters, e.g. \(\mathcal{M}\), denote sets. \(\abs{\mathcal{M}}\) denotes the cardinality of  set $\mathcal{M}$. The indicator function $ \idc{\mathcal{A}}$ is equal to $1$ if the event \(\mathcal{A}\) is true, and $0$ otherwise. Let $X$ and $Z$ be respectively  an input and output  to a channel $P_{Z|X}$, which is a (possibly stochastic) mapping from the input alphabet $\mathcal{X}$ to the output alphabet $\mathcal{Z}$.
The mutual information \(I(X;Z)\) is denoted by \(I(P_X, P_{Z|X})\). 
The Bernoulli distribution with parameter $p$ is denoted by $\bern(p)$ and the binary symmetric channel with parameter $q$ is denoted by $\bsc(q)$. 
For  $p,q \in [0,1]$, we define $ p \ast q \triangleq (1-q)p + q(1-p) $.
\section{Problem Setting}
\label{sec:problem_setting}
We consider the setting introduced in Section \ref{sec:introduction} and illustrated in Fig. \ref{fig:block_diagram} with finite alphabets $\mathcal{X}$, $\mathcal{Z}$, $\mathcal{Y}$ and a binary parameter $\theta \in \{0,1 \}$.
The channels are  stationary and memoryless, that is
\begin{equation}
\label{eq:DMCs}
P_{Z^n|X^n}( z^n | x^n)  = \prod_{i = 1}^{n}  P_{Z|X}( z_i | x_i) \quad \text{and}
\quad   P_{Y^n|X^n}^{\theta}( y^n | x^n)  = \prod_{i = 1}^{n}  P_{Y|X}^{\theta}( y_i | x_i).
\end{equation}
An admissible input sequences $x^n \in \mathcal{X}^n$ must satisfy an average cost constraint of 
\begin{equation}
\label{eq:cost_constraint}
    \frac{1}{n} \sum_{i = 1}^{n} b(x_i) \leq B
\end{equation}
where  $b:\mathcal{X} \to \mathbb{R}_{+}$ is some non-negative cost function and $B \geq 0$ is the average cost constraint.

To simplify the notation in what follows, we use $W_n(y^{n}|x^{n}) = \prod_{i=1}^{n}W(y_i | x_i)$  and $V_n(y^{n}|x^{n}) = \prod_{i=1}^{n}V(y_i | x_i)$  to denote $P_{Y^{n} | X^{n}}^{0}(y^{n} | x^{n})$ and  $P_{Y^{n} | X^{n}}^{1}(y^{n} | x^{n})$ respectively,  where $W$ and $V$ respectively denote $P_{Y|X}^0$ and $P_{Y|X}^1$. We assume that \(W(y|x)V(y|x) \neq 0\), for every $x \in \mathcal{X}$ and  $y \in \mathcal{Y}$, which holds for most channels of interest.
\subsection{Codes and error probabilities}
\label{subsec:codes}
For any positive integers  $n$ and $\abs{\mathcal{M}_n}$, an $(n,\abs{\mathcal{M}_n})$-code for the above setting
consists of a message set given by $\mathcal{M}_n \triangleq \big\{1,2,\ldots, \abs{\mathcal{M}_n} \big\}$ and the following mappings:
\begin{itemize}
\item An encoding function $f_n : \mathcal{M}_n \to \mathcal{X}^n$ that maps each message $m \in \mathcal{M}_n$ into a  codeword  $x^n = f_n(m)$ that satisfies the cost constraint in \eqref{eq:cost_constraint}.
The corresponding codebook $\mathcal{C}_n$ is the set of all $|\mathcal{M}_n|$ codewords.
\item A message decoding function  $\varphi_n : \mathcal{Z}^n \to \hat{\mathcal{M}}_n$ that maps each output sequence $z^n \in \mathcal{Z}^n $  into  a decoded message $\hat{m} = \varphi_n(z^n)$ in $\hat{\mathcal{M}}_n$, where 
we may assume that $ \hat{\mathcal{M}}_n = \mathcal{M}_n$.
\item A channel discrimination function $\psi_n : \mathcal{Y}^n \times \mathcal{M}_n \to \hat{\Theta} $ that maps each output sequence and message pair $(y^n,m) \in \mathcal{Y}^n \times \mathcal{M}_n$ into a decision  $\hat{\theta} = \psi_n(y^n,m)  $ in $\hat{\Theta}$, where $\hat{\Theta}$ is set to $\{0,1\}$.
\end{itemize}

The message $M$, which is drawn randomly from $\mathcal{M}_n $, is encoded into $X^n = f_n(M)$ and then sent over the channels. 
Upon observing $Z^n$, the receiver produces a decoded message $\hat{M} = \varphi_n(Z^n)$.
On the other hand, upon observing $Y^n$ and with knowledge of $M$, the sensor produces a binary decision $\hat{\theta} = \psi_n(Y^n,M)$.
The discrimination function depends on the message $M$ only through the codeword $X^n = f_n(M)$. 
Therefore we will often write $\psi_n(Y^n,X^n)$ instead of $\psi_n(Y^n,M)$ henceforth. 

\emph{Decoding error:} For a given code, the probability of decoding error given that message $M = m$ has been sent is $\Prb\left[ \varphi_n(Z^n) \neq m \mid M =m \right]$.
The maximum probability of decoding error is defined as 
\begin{IEEEeqnarray}{C}
    P_{\mathrm{e},n} \triangleq \max_{m \in \mathcal{M}_n} \Prb\left[ \varphi_n(Z^n) \neq m \mid M =m \right]
\end{IEEEeqnarray}
which is a common performance measure that reflects the assumption that messages are equally important.

\emph{Discrimination error:}  
There are two types of discrimination errors associated with the two values of $\theta$.
Given $M = m$, and hence the codeword $f_n(m) = x^n \in \mathcal{C}^n$ is sent, the two types of error probability are defined as 
\begin{align}
\label{eq:discrimination_error_probabilities}
\varepsilon_{0,n}(x^n) & \triangleq   \Prb \left[ \psi_n(Y^n,X^n) \neq \theta  \mid  \theta = 0, X^n = x^n  \right] \\
\varepsilon_{1,n}(x^n)  & \triangleq  \Prb \left[ \psi_n(Y^n,X^n) \neq \theta \mid   \theta = 1, X^n = x^n  \right]
\end{align}
known respectively as the type I and type II error.
We treat the two types equally, and hence $\psi_n$ is designed to minimize the worst of the two. This yields a discrimination error, given $X^n = x^n$, of
\begin{equation}
\varepsilon_{n}(x^n) \triangleq  \min_{\psi_n(\cdot, x^n)} \max \left\{ \varepsilon_{0,n}(x^n), \varepsilon_{1,n}(x^n) \right\}
\end{equation}
where $\psi_n(\cdot, x^n)$ is designed with knowledge of $x^n$.
Since it is not known beforehand which codeword in $\mathcal{C}^n$ will be sent, it is reasonable to define the 
discrimination error by taking the maximum over codewords in $\mathcal{C}^n$ as 
\begin{equation}
\varepsilon_{n} \triangleq  \max_{x^n \in \mathcal{C}_n} \varepsilon_{n}(x^n).
\end{equation}
By considering $\varepsilon_{n}$, a certain error performance is guaranteed regardless of which messages has been selected. 

For some applications, it may be desirable to treat the two types of discrimination error unequally. This will be addressed in Section \ref{sec:Neyman_Pearson}, where we adopt a Neyman-Pearson criterion for channel discrimination.
\subsection{Rate-Exponent region}
We are interested in the asymptotic performance limits measured in terms of the message communication rate and the channel discrimination error exponent. 
They are formalized as follows.
\begin{definition}
\label{definition:R_E_region}
A rate-exponent tuple $(R,E)$ is said to be achievable if there exists a sequence of 
$\left(n,|\mathcal{M}_n|\right)$-codes as defined in Section \ref{subsec:codes} such that $ \lim_{n \to \infty} P_{\mathrm{e},n} = 0$ and 
\begin{equation}
\nonumber
R = \liminf_{n \to \infty} \frac{1}{n} \log  |\mathcal{M}_n|  \quad \text{and} \quad 
E = \liminf_{n \to \infty} \frac{1}{n} \log \frac{1}{\varepsilon_{n}}.
\end{equation}
The rate-exponent region $\mathcal{R}$ is the closure of the set of all achievable pairs $(R,E)$.
\end{definition}
The main result of this paper is a characterization of the rate-exponent region $\mathcal{R}$ for the general discrete memoryless channels in \eqref{eq:DMCs} under the average input cost constraint in \eqref{eq:cost_constraint}.
\subsection{Types and typical sequences}
Here we present some notation and preliminaries on types and strongly typical sequences from \cite{csiszarInformationTheoryCoding2011}, which are essential for the statement and proofs of our results.
Given a sequence  \(x^{n} \in \mathcal{X}^{n}\), we define
\begin{IEEEeqnarray}{C}
    N(a|x^{n}) \triangleq \sum_{i=1}^{n}\idc{x_i = a}, a \in \mathcal{X}.
\end{IEEEeqnarray}
The type of \(x^{n}\), denoted by \(\mathsf{P}_{x^{n}}\), is  a distribution on \(\mathcal{X}\) defined as
\begin{IEEEeqnarray}{C}
    \mathsf{P}_{x^{n}} (a)= \frac{N(a|x^{n})}{n}, \ a \in \mathcal{X}.
\end{IEEEeqnarray}
Let \(\mathcal{P}(\mathcal{X})\) be the set of all distributions (i.e. probability mass functions) on $\mathcal{X}$ and \(\mathcal{P}_n(\mathcal{X})\) be the set of all types of sequences in $\mathcal{X}^n$. Note that \(\mathcal{P}_n(\mathcal{X}) \subset \mathcal{P}(\mathcal{X})\). Moreover, the number of types in \(\mathcal{P}_n(\mathcal{X}) \) is bounded as follows
\begin{equation}
\label{eq:type_counting_bound}
\abs{\mathcal{P}_n(\mathcal{X})}  \leq (n+1)^{\abs{\mathcal{X}}}
\end{equation}
see, e.g., \cite[Lemma 2.2]{csiszarInformationTheoryCoding2011}.
An arbitrary member of \(\mathcal{P}(\mathcal{X})\) is denoted by $P_X$ or $P$, and a member of 
\(\mathcal{P}_n(\mathcal{X})\) is denoted by \(\mathsf{P}_{x^{n}}\) or \(\mathsf{P}\) for emphasis.
A sequence \(x^{n} \in \mathcal{X}^n\) is called \(P\)-typical with constant \(\delta_n\) if
\begin{equation}
\abs{ \mathsf{P}_{x^{n}}(a)- P(a)}  \leq \delta_n, \ a \in \mathcal{X}    
\end{equation}
with the additional condition that $\mathsf{P}_{x^{n}}(a) = 0$ whenever $ P(a) = 0$.
This notion of typicality is also called strong typicality, as opposed to the notion of weak typicality (or entropy typicality)  \cite{coverElementsInformationTheory2006}.

The set of all sequences \(x^{n} \in \mathcal{X}^n\) that are \(P\)-typical with constant \(\delta_n\) is denoted by \(\mathcal{T}^{n}_{[P]_{\delta_n}}\).
We adopt the common convention that \(\delta_n\) satisfies \(\delta_n \to 0\) and \(\sqrt{n}\delta_n \to \infty\) when \(n \to \infty\) \cite[Convention 2.11]{csiszarInformationTheoryCoding2011}.
We will drop \(\delta_n\) and simply write \(\mathcal{T}_{[P]}^{n}\) henceforth, where it is understood that the above convention holds.
\section{Channel Discrimination Exponent}
We first consider the channel discrimination problem for a given codebook, and in doing so we review and adapt known results on binary hypothesis testing, which will be useful in proving the main result in the next section. 

Let $\mathcal{C}_n$ be an arbitrary codebook and suppose that a codeword $x^n \in \mathcal{C}_n$ has been sent by the transmitter.
With knowledge of $x^n$, channel discrimination boils down to simple hypothesis testing of $W_n( y^n | x^n)$ versus  $V_n( y^n | x^n)$.
Given a sequence of codebooks $(\mathcal{C}_n)_{n \in \mathbb{N}}$, the best channel discrimination exponent under the maximum error criterion is characterized in the following result. 
For stating this result, we define
\begin{IEEEeqnarray}{C}
\label{eq:def_C}
    C(W\|V | P_X)=  - \min_{s \in [0,1]} \sum_{x \in \mathcal{X}} P_X(x) \log \left( \sum_{y \in \mathcal{Y}} W(y|x)^{1-s}V(y|x)^{s} \right).
\label{eq:disc_no_cost_E0}
\end{IEEEeqnarray}
\begin{theorem}
\label{theorem:optimum_E_given_codebook}
Given a sequence codebooks $(\mathcal{C}_n)_{n \in \mathbb{N}}$, the channel discrimination exponent is given by
\begin{equation}
\label{eq:optimum_E_given_codebook}
    E = \liminf_{n \to \infty} \min_{x^n \in \mathcal{C}_n}  C(W\|V | \mathsf{P}_{x^n}).
\end{equation}
\end{theorem}
Next we present a proof for the above theorem. 
\subsection{Proof of Theorem \ref{theorem:optimum_E_given_codebook}}
Any deterministic discrimination function $\psi_n$, or hypothesis test, is characterized by a decision region $\Omega \subseteq \mathcal{Y}^n$, where $\psi_n(y^n,x^n) = 0$ when $y^n \in \Omega$ and $\psi_n(y^n,x^n) = 1$ otherwise. 
A decision region of interest is the one resulting from the likelihood ratio test, given for some threshold $\tau \in \mathbb{R}$ by
\begin{IEEEeqnarray}{rCl}
\Omega(\tau) &=& \left \{ y^{n} \in \mathcal{Y}^{n} : \log \frac{W_n(y^{n}|x^n)}{V_n(y^{n}|x^n)} \geq \tau \right \}. \label{eq:pre_LRT_region1}
\end{IEEEeqnarray}
\subsubsection{Achievability} Given that $x^n$ is sent, use the likelihood ratio test with $\tau = 0$. A well known upper bound is
\begin{align}
\varepsilon_{0,n}(x^n) & =  \sum_{y^n \in \mathcal{Y}^n} W_n(y^n|x^n)  \idc{W_n(y^n|x^n) < V_n(y^n|x^n)} \\
\label{eq:error_typeI_UB}
& \leq \sum_{y^n \in \mathcal{Y}^n} W_n(y^n|x^n) \left( \frac{V_n(y^n|x^n)}{W_n(y^n|x^n)} \right)^{s}
\end{align}
which holds for any parameter $s \in [0,1]$. The exact same upper bound also holds for  $\varepsilon_{1,n}(x^n) $. 
By taking the logarithm of the quantity in \eqref{eq:error_typeI_UB}, we define the following function of the $s$ parameter
\begin{IEEEeqnarray}{C}
\mu(s|x^n)  \triangleq  \log \left( \sum_{y^{n} \in \mathcal{Y}^{n}}  W_n(y^{n} | x^n)^{1-s}V_n(y^{n} | x^n)^{s}\right).
\end{IEEEeqnarray}
\(\mu(s|x^n)\) is strictly convex on \(s \in [0,1]\) (see the proof of \cite[Theorem 5]{shannonLowerBoundsError1967}).
Let \(s_0\) minimize $\mu(s|x^n)$ on $s\in [0,1]$. It follows that the two types of error probability are bounded above as
\begin{equation}
\label{eq:Chernoff_bound}
\max \left\{ \varepsilon_{0,n}(x^n), \varepsilon_{1,n}(x^n) \right\} \leq e^{\mu(s_0|x^n)}.
\end{equation}
This is known as the Chernoff information bound \cite[Section 11.9]{coverElementsInformationTheory2006}, and $-\mu(s_0|x^n)$ is equal to the Chernoff information between $W_n( \cdot | x^n)$ and $V_n( \cdot | x^n)$, denoted by $C\big( W_n( \cdot | x^n) \| V_n( \cdot | x^n) \big)$.
This can be expressed as 
\begin{IEEEeqnarray}{rCl}
C\big( W_n( \cdot | x^n) \| V_n( \cdot | x^n) \big)  & \triangleq & - \log \left( \sum_{y^{n} \in \mathcal{Y}^{n}}  W_n(y^{n} | x^n)^{1-s_0}V_n(y^{n} | x^n)^{s_0}\right) \\
\label{eq:memoryless_mu}
& =& - n \sum_{x \in \mathcal{X}} \mathsf{P}_{x^n}(x) \log \left( \sum_{y \in \mathcal{Y}} W(y|x)^{1-s_0}V(y|x)^{s_0} \right) \\
& =& n C(W\|V | \mathsf{P}_{x^n})
\end{IEEEeqnarray}
where \eqref{eq:memoryless_mu} is obtained using the fact that $W_n$ and $V_n$ are memoryless. 
Hence \eqref{eq:Chernoff_bound} can be expressed as
\begin{equation}
\label{eq:Chernoff_bound_2}
\max \left\{ \varepsilon_{0,n}(x^n), \varepsilon_{1,n}(x^n)  \right\}  \leq e^{- n C(W\|V | \mathsf{P}_{x^n}) }
\end{equation}
Under the chosen test $\Omega(0)$, let $ x^n(1) \in \mathcal{C}^n$ be such that
\begin{IEEEeqnarray}{C}
    x^n(1) = \arg\max_{x^n \in \mathcal{C}_n} \max \left\{ \varepsilon_{0,n}(x^n), \varepsilon_{1,n}(x^n)  \right\}
\end{IEEEeqnarray}
where, without loss of generality,  $x^n(1)$ is the first codeword in $\mathcal{C}^n$. Then
\begin{align}
E & \triangleq \liminf_{n \to \infty} -\frac{1}{n} \log \max_{x^n \in \mathcal{C}_n} \min_{\psi_n(\cdot, x^n)} \max \left\{ \varepsilon_{0,n}(x^n), \varepsilon_{1,n}(x^n)  \right\}
\\
& \geq \liminf_{n \to \infty} -\frac{1}{n} \log \max_{x^n \in \mathcal{C}_n} \max \left\{ \varepsilon_{0,n}(x^n), \varepsilon_{1,n}(x^n)  \right\} \label{eq:det_test}\\
& = \liminf_{n \to \infty} -\frac{1}{n} \log  \max \left\{ \varepsilon_{0,n}(x^n(1)), \varepsilon_{1,n}(x^n(1))  \right\}\\
&\geq  \liminf_{n \to \infty}  C(W\|V | \mathsf{P}_{x^n(1)}) \\
& \geq  \liminf_{n \to \infty}  \min_{x^n \in \mathcal{C}_n} C(W\|V | \mathsf{P}_{x^n})
\end{align}
where in \eqref{eq:det_test}, we use the chosen test $\Omega(0)$ (which may not be optimal). This completes the achievability part.
\subsubsection{Converse} This is obtained from the following lower bound due to Shannon, Gallager and Berlekamp \cite{shannonLowerBoundsError1967}.
\begin{theorem*}
\textbf{(Corollary of \cite[Theorem 5]{shannonLowerBoundsError1967}).}
\label{thm:pre_up_low_bounds}
For any decision region \(\Omega \subseteq \mathcal{Y}^n\), at least one of the following holds
\begin{IEEEeqnarray}{rCl}
\label{eq:error_LB_1}
\varepsilon_{0,n}(x^n) &\geq& \frac{1}{4} e^{\mu(s_0|x^n)-s_0 \sqrt{2\mu^{\prime \prime}(s_0|x^n)}} \\
\label{eq:error_LB_2}
\varepsilon_{1,n}(x^n) &\geq& \frac{1}{4} e^{\mu(s_0|x^n) - (1-s_0) \sqrt{2\mu^{\prime \prime}(s_0|x^n)}}
\end{IEEEeqnarray}
where $\mu^{\prime \prime}(s_0|x^n)$ is the second derivative of $\mu(s|x^n)$ at $s = s_0$, and  $\sqrt{2\mu^{\prime \prime}(s_0|x^n)}$ is proportional to $\sqrt{n}$.
\end{theorem*}
 The above lower bounds hold for deterministic discrimination functions (i.e. deterministic tests), while the optimal test for minimizing the maximum error probability may not be deterministic.
We resolve this issue by using the Bayesian error to bound the maximum error, as seen below. To this end, let $x^n(1)$ achieve
\begin{IEEEeqnarray}{C}
    C(W\|V | \mathsf{P}_{x^n(1)}) = \min_{x^n \in \mathcal{C}_n}  C(W\|V | \mathsf{P}_{x^n}).
\end{IEEEeqnarray}
It follows that
\begin{align}
\label{eq:E_UB_fixed_codebook}
E & =\liminf_{n \to \infty} -\frac{1}{n} \log \max_{x^n \in \mathcal{C}_n} \min_{\psi_n(\cdot, x^n)} \max \left\{ \varepsilon_{0,n}(x^n), \varepsilon_{1,n}(x^n)  \right\} \\
  & \leq \liminf_{n \to \infty} -\frac{1}{n} \log   \max_{x^n \in \mathcal{C}_n} \min_{\psi_n(\cdot, x^n)} \left\{ 0.5\varepsilon_{0,n}(x^n) + 0.5\varepsilon_{1,n}(x^n) \right\} \label{eq:bayesian_error}\\
& \leq \liminf_{n \to \infty} -\frac{1}{n} \log \min_{\psi_n(\cdot, x^n)}  \left\{ 0.5\varepsilon_{0,n}(x^n(1))+0.5 \varepsilon_{1,n}(x^n(1))  \right\} \\
 & \leq \liminf_{n \to \infty}  -\frac{1}{n}  \left( \mu(s_0|x^n(1)) - \sqrt{2\mu^{\prime \prime}(s_0|x^n(1))} \right) \label{eq:bayesian_optimal_test}\\
& = \liminf_{n \to \infty} \min_{x^n \in \mathcal{C}_n}  C(W\|V | \mathsf{P}_{x^n}).
\end{align}
In \eqref{eq:bayesian_error}, we transition from the maximum error to the Bayesian error, for which the optimal test is deterministic and  known as the \emph{maximum a posteriori} test,  see \cite[Section 11.9]{coverElementsInformationTheory2006}.  Therefore, from \eqref{eq:bayesian_error} onward, we can limit our attention to deterministic tests and employ the lower bounds in \eqref{eq:error_LB_1} and \eqref{eq:error_LB_2} without loss of generality.
We use these bounds, alongside $\max\{s_0,(1-s_0)\} \leq 1$, to obtain \eqref{eq:bayesian_optimal_test}. This completes the proof of Theorem \ref{theorem:optimum_E_given_codebook}.
\subsection{Best discrimination exponent}
The next corollary follows directly from Theorem \ref{theorem:optimum_E_given_codebook}.
\begin{corollary}
\label{corollary:best_exponent}
Let $E^{\star} \triangleq \max_{(R,E) \in \mathcal{R}} E$ denote the best possible channel discrimination exponent. Then
\begin{equation}
\label{eq:optimal_max_exponent}
    E^{\star} = \max_{P_X: \E_{P_X}[b(X)] \leq B}  C(W\|V | P_X)
\end{equation}
\end{corollary}
\begin{IEEEproof}
Let  \(P_X^{\star}\) be a distribution that attains the maximum in \eqref{eq:optimal_max_exponent}.
From \eqref{eq:optimum_E_given_codebook}, we have $E^{\star} \leq C(W\|V | |P_X^{\star})$, 
which holds for any sequence of codebooks.
This upper bound is achievable as follows. Let $(\mathcal{C}_n)_{n \in \mathbb{N}}$ be a sequence of single-codeword codebooks such that $\mathcal{C}_n = \{ x^n\}$ and $\mathsf{P}_{x^n} \to P_X^{\star}$ as $n \to \infty$. In this case we have
\begin{align}
E  = \liminf_{n \to \infty}  C(W\|V | \mathsf{P}_{x^n})  =  C(W\|V | P_X^{\star})
\end{align}
where the last equality is by continuity of $C(W\|V | P_X)$ in $P_X$  as shown in Appendix \ref{appendix:Continuity_of_C}.
\end{IEEEproof}
\section{Communication-Discrimination Trade-off}
\label{sec:comm_disc_trade_off}
Here we present the main result, where we characterize the
rate-exponent region introduced in Definition \ref{definition:R_E_region}.
\begin{theorem}
\label{thm:disc_cost}
\(\mathcal{R}\) is given by the set of all non-negative pairs \((R,E)\) such that 
\begin{IEEEeqnarray}{Cl}
&R \leq I(P_X, P_{Z|X}) \\
&E \leq C(W\|V | P_X)
\end{IEEEeqnarray}
for some input distribution \(P_X\) on \(\mathcal{X}\) that satisfies \(\E_{P_X}[b(X)] \leq B\).
\end{theorem}
The proof of the above theorem is presented in the next subsection. We can obtain the following equivalent representation of $\mathcal{R}$ from the converse proof of Theorem \ref{thm:disc_cost}.
\begin{corollary}
\label{corollary:equiv_region}
\(\mathcal{R}\) in Theorem \ref{thm:disc_cost} is equivalently characterized by all non-negative pairs \((R,E)\) such that
\begin{align}
\label{eq:equiv_region_E}
    E & \leq  \max_{P_X: \E_{P_X}[b(X)] \leq B } C(W\|V | P_X) \\
\label{eq:equiv_region_R}    
    R & \leq \max_{P_X: C(W\|V | P_X) \geq E, \ \E_{P_X}[b(X)] \leq B} I(P_X, P_{Z|X}).
\end{align}
\end{corollary}
Next, we present a couple of examples to illustrate the result in Theorem \ref{thm:disc_cost}.
\begin{example}
\label{example:1}
Consider a setting with a binary input, and binary outputs given by
\begin{equation}
Z  = X \oplus N_{Z}  \ \text{and} \ Y  = \theta X \oplus N_{Y}  
\end{equation}
where $N_{Z}$ and $N_{Y}$ are Bernoulli with parameters $p$ and $q$, respectively.
Here $P_{Z|X}$ is a $\bsc(p)$, $V$ is a $\bsc(q)$, while $W$ satisfies $W(1|\cdot) = q$ and $W(0|\cdot)=1-q$.
Let $P_X \sim \bern (\rho)$, then
\begin{IEEEeqnarray}{rcl}
    I(P_X, P_{Z|X}) &=& H(\rho * p) -H(p), \\
    C(W\|V|P_X) &=& - \min_{s \in [0,1]} \rho \log \left( (1-q)^{1-s}q^s + q^{1-s}(1-q)^{s} \right) \label{eq:s_0.5}.
\end{IEEEeqnarray}
Let $g(s) =  \log \left( (1-q)^{1-s}q^s + q^s(1-q)^{1-s} \right)$, which is strictly convex in $s$ (recall that $\mu(s|x^n)$ is strictly convex). 
Since $g(s) = g(1-s)$, it follows that $s=0.5$ is the  minimizer in \eqref{eq:s_0.5}. Consequently,
\begin{IEEEeqnarray}{c}
    C(W\|V|P_X) = -\rho \log (2\sqrt{(1-q)q}) = -\rho \log e^{-D(0.5\| q)} = \rho D(0.5\| q).
\end{IEEEeqnarray}
Therefore, $\mathcal{R}$ in this case is described by
\begin{IEEEeqnarray}{cl}
    & R \leq H(\rho * p) - H(p) \\
    & E \leq \rho D(0.5\| q)
\end{IEEEeqnarray}
for some $\rho \leq B$, which is the Bernoulli parameter of $P_X$ (we assume  $b(1) = 1$ and $B \leq 1$).
Here the maximum rate is achieved when $\rho= \min\{0.5,B\}$, while the maximum exponent is achieved when $\rho = B$. Hence, there is a trade-off between the rate and the exponent whenever $B > 0.5$.
\end{example}
\begin{example}
\label{example:2}
Consider a binary input binary output settings as in the previous example, but here  $W$ is a $\bsc(p)$ and $V$ is a $\bsc(q)$. In this case we have
\begin{IEEEeqnarray}{c}
    C(W\|V|P_X) = -\min_{s \in [0,1]} \log \left( (1-p)^{1-s}(1-q)^{1-s} +  p^sq^s\right).
\end{IEEEeqnarray}
and the exponent inequality $E \leq C(W\|V|P_X)$ does not depend on the input cost constraint, which only affects the rate $R$. 
Hence there is no trade-off between $R$ and $E$ here, and $\mathcal{R}$ is a rectangle.
\end{example}
\subsection{Proof of Theorem  \ref{thm:disc_cost} } 
\label{subsec:proof_theorem_trade_off}
To prove Theorem  \ref{thm:disc_cost}, we rely on the exponent characterization for a given sequence of codebooks in Theorem \ref{theorem:optimum_E_given_codebook} and combine it with a coding argument for discrete memoryless channels with a cost constraint  \cite[Theorem 6.11]{csiszarInformationTheoryCoding2011}. 
\subsubsection{Achievability} Let  \(P_X\) be an arbitrary input distribution on $\mathcal{X}$ which satisfies \(\E_{P_X}[b(X)] \leq B - \delta\) for some $\delta > 0$. We make use of the following achievability result, borrowed from Csisz{\'a}r and K{\"o}rner \cite{csiszarInformationTheoryCoding2011}. 
\begin{lemma}
\textbf{\cite[Corollary 6.3]{csiszarInformationTheoryCoding2011}.}
\label{lm:pre_M_low_bound}
For any discrete memoryless channel \(P_{Z|X}: \mathcal{X} \to \mathcal{Z}\) and any input distribution \(P_X\) on \(\mathcal{X}\), there exists a sequence of \((n,\abs{\mathcal{M}_n})\)-codes with \(P_X\)-typical codebooks, i.e.  $\mathcal{C}_n \subseteq \mathcal{T}^{n}_{[P_X]}$ for all $n$,  such that  \( \liminf_{n \to \infty} \frac{1}{n} \log \abs{\mathcal{M}_n}   \geq   I(P_X, P_{Z|X}) \) and \(\lim_{n \to \infty} P_{\mathrm{e},n} = 0 \).
\end{lemma}
Using a sequence of codes from Lemma \ref{lm:pre_M_low_bound}, we guarantee that the rate \(R = I(P_X, P_{Z|X})\) is achieved.
Moreover, since any codeword $x^{n} \in \mathcal{C}^n$ is also in \(\mathcal{T}_{[P_X]}^{n}\), it follows that
\begin{align}
 \frac{1}{n}\sum_{i=1}^{n}b(x_i) & = \E_{\mathsf{P}_{x^{n}}}[b(X)] \leq \E_{P_X}[b(X)] + \delta_n \sum_{x\in \mathcal{X}}b(x)  \leq B - \delta'_n
\end{align}
where $\delta'_n  \triangleq \delta  - \delta_n \sum_{x\in \mathcal{X}}b(x)$, which can be made as small as desired provided that $n$ is large enough. Therefore, the cost constraint is also satisfied. It remains to show that  \(E = C(W\|V | P_X)\) is achieved. 

For any $n \in \mathbb{N}$, assume that
\begin{equation}
x^n(1)= \arg\min_{x^n \in \mathcal{C}^n} C(W\|V | \mathsf{P}_{x^n}).
\end{equation}
From Theorem \ref{theorem:optimum_E_given_codebook}, we know that the following exponent is achievable
\begin{align}
     E & = \liminf_{n \to \infty} \min_{x^n \in \mathcal{C}_n}  C(W\|V | \mathsf{P}_{x^n}) \\
     & = \liminf_{n \to \infty}  C(W\|V | \mathsf{P}_{x^n(1)}).
\end{align}
Since $x^n(1) \in \mathcal{T}_{[P_X]}^{n}$ for all $n$, then 
$\mathsf{P}_{x^n(1)} \to P_X$ as $n \to \infty$, and by continuity we have
\begin{align}
    E =   C(W\|V | P_{X}).
\end{align}
\subsubsection{Converse} Suppose that we have a sequence of $\left(n,|\mathcal{M}_n| \right)$-codes such that $ \lim_{n \to \infty}P_{\mathrm{e},n} = 0$, and let $(\mathcal{C}^n)_{n \in \mathbb{N}}$ be the corresponding sequence of codebooks.
From the converse of Theorem \ref{theorem:optimum_E_given_codebook}, we know that we must have 
\begin{align}
\label{eq:E_UB_converse}
E & \leq \liminf_{n \to \infty}  \min_{x^n \in \mathcal{C}_n}C(W\|V | \mathsf{P}_{x^n}).
\end{align}
Now we wish to find an upper bound on the number of codewords in $\mathcal{C}^n$ as $n$ grows large.
To this end, we make use of the following result, which we also borrow from \cite{csiszarInformationTheoryCoding2011}.
\begin{lemma}
\textbf{\cite[Corollary 6.4]{csiszarInformationTheoryCoding2011}.}
\label{lm:pre_M_up_bound}
Consider any sequence of \((\mathcal{M}_n, n)\)-codes for the channel $P_{Z|X} $ with \(P_{\mathrm{e},n} = \epsilon\) and $P_{X}$-typical codebooks $(\mathcal{C}_n)_{n \in \mathbb{N}}$, i.e.  $\mathcal{C}_n \subseteq \mathcal{T}^{n}_{[P_X]}$ for all $n$.
For any  $\epsilon,\tau \in (0,1)$, there exists a sufficiently large $ n_{\tau, \epsilon}$, which depends on $\tau$ and $\epsilon$, such that for all $n \geq n_{\tau,\epsilon}$ we have
\begin{IEEEeqnarray}{C}
    \frac{1}{n}\log \abs{\mathcal{M}_n} < I(P_X, P_{Z|X}) + \tau.
\end{IEEEeqnarray}
\end{lemma}
Any codebook $\mathcal{C}^n$ can be partitioned into subsets, where codewords in the same subset have the same type.
In a subset with type $\mathsf{P}$, Lemma \ref{lm:pre_M_up_bound} suggest that the number of codewords 
is at most $\exp\left(nI(\mathsf{P}, P_{Z|X}) + n\tau \right)$, for large enough $n$.  
It follows that for any codebook $\mathcal{C}^n$ with \(P_{\mathrm{e},n} = \epsilon\) and for $n \geq n_{\tau,\epsilon}$, we have the upper bound 
\begin{equation}
\label{eq:message_set_UB_converse}
 \abs{\mathcal{M}_n}  \leq \sum_{\mathsf{P} \in \{\mathsf{P}_{x^n}:x^n \in \mathcal{C}^n\}} e^{nI(\mathsf{P}, P_{Z|X}) + n\tau}
\end{equation}
where the summation is over the distinct codeword types.
We now find conditions on admissible codeword types.

The upper bound in \eqref{eq:E_UB_converse} implies that for any $\delta > 0$, there exists a large enough $n_{\delta}$ such that
\begin{equation}
\label{eq:E_UB_converse_2}
 \inf_{n\geq n_{\delta}} \min_{x^n \in \mathcal{C}_n} C(W\|V | \mathsf{P}_{x^n}) \geq E - \delta.
\end{equation}
Let $\mathcal{P}_{E-\delta}(\mathcal{X})$ be a subset of distributions in $\mathcal{P}(\mathcal{X})$ defined as
\begin{IEEEeqnarray}{C}
    \mathcal{P}_{E-\delta}(\mathcal{X}) \triangleq \left\{ P_X : C(W\|V | P_X) \geq E - \delta, \ \E_{P_X}[b(X)] \leq B  \right\}
\end{IEEEeqnarray}
and let $P_{X}^{\star}$ be a distribution in $\mathcal{P}_{E-\delta}(\mathcal{X}) $ that maximizes $I(P_X, P_{Z|X})$.
It follows from \eqref{eq:E_UB_converse_2} that for all $n \geq n_{\delta}$, we must have $\mathsf{P}_{x^n} \in \mathcal{P}_{E-\delta}$ for every $x^n \in \mathcal{C}_n$.
Combing this with \eqref{eq:message_set_UB_converse}, and taking $n \geq \max\{n_{\tau,\epsilon},n_{\delta}\}$, we get
\begin{align}
\abs{\mathcal{M}_n} & \leq \sum_{\mathsf{P} \in \mathcal{P}_{E - \delta}(\mathcal{X}) } e^{nI(\mathsf{P}, P_{Z|X}) + n\tau} \\
 & \leq | \mathcal{P}_{E - \delta} | e^{nI(P_X^{\star}, P_{Z|X}) + n\tau} \\
 & \leq (n+1)^{|\mathcal{X}|} e^{nI(P_X^{\star}, P_{Z|X}) + n\tau}
\end{align}
where the last inequality follows from the type counting bound in \eqref{eq:type_counting_bound}.
For large enough $n$, we have
\begin{equation}
\frac{1}{n} \log \abs{\mathcal{M}_n} \leq I(P_X^{\star}, P_{Z|X}) + \tau + \frac{|\mathcal{X}|}{n} \log(n+1).
\end{equation}
By taking $n \to \infty$ and  $\delta,\tau,\epsilon \to 0$, we obtain 
\begin{equation}
\liminf_{n \to \infty}\frac{1}{n} \log \abs{\mathcal{M}_n} \leq I(P_X^{\star}, P_{Z|X}) = \max_{P_X: C(W\|V | P_X) \geq E , \ \E_{P_X}[b(X)] \leq B } I(P_X, P_{Z|X}).
\end{equation}
Moreover, from Corollary \ref{corollary:best_exponent}, we know that
\begin{equation}
    E \leq  \max_{P_X: \E_{P_X}[b(X)] \leq B}  C(W\|V | P_X).
\end{equation}
Therefore, $(R,E)$ must be in the region described in Corollary \ref{corollary:equiv_region}. 
Note that this region is also obviously achievable through the argument of achievability, for example, choose $P_X$ to be $P_X^{\star}$. This concludes the proof. 

\section{Trade-off under Neyman-Pearson Channel Discrimination}
\label{sec:Neyman_Pearson}
Here we consider the case where the two types of discrimination errors are treated unequally.
We adopt the Neyman-Pearson criterion, where the focus is on minimizing one type of error while keeping the other under control.
Here we choose to minimize the type II error probability while requiring that the type I error probability does not exceed a desired threshold $\alpha  \in (0,1)$. For a codebook $\mathcal{C}^n$ and given that the codeword $x^n$ has been sent, $\psi_n$ is designed according to the above criterion, and the resulting type II discrimination error is given by
\begin{equation}
\beta_{\alpha,n}(x^n) \triangleq  \min_{\psi_n(\cdot, x^n) : \varepsilon_{0,n}(x^n) \leq \alpha}  \varepsilon_{1,n}(x^n).
\end{equation}
As argued in Section \ref{subsec:codes}, since it is not known beforehand which codeword in $\mathcal{C}^n$ will be sent, we take the maximum over all codewords in $\mathcal{C}^n$ and obtain an error probability  of
\begin{equation}
\beta_{\alpha,n} \triangleq  \max_{x^n \in \mathcal{C}^n} \beta_{\alpha,n}(x^n).
\end{equation}

\subsection{Asymptotic Trade-off}
Under the Neyman-Pearson criterion,  the asymptotic trade-off is formalized as follows.
\begin{definition}
Under the Neyman-Pearson discrimination criterion, the rate-exponent tuple $(R,E_{\alpha})$ 
is  achievable if there exists a sequence of 
$\left(n,|\mathcal{M}_n|\right)$-codes such that $ \lim_{n \to \infty} P_{\mathrm{e},n} = 0$ and
\begin{equation}
\nonumber
R = \liminf_{n \to \infty} \frac{1}{n} \log  |\mathcal{M}_n| 
\quad \text{and} \quad
E_{\alpha} = \liminf_{n \to \infty} \frac{1}{n} \log \frac{1}{\beta_{\alpha,n}}.
\end{equation}
The rate-exponent region $\mathcal{R}_{\alpha}$ is the closure of the set of all achievable pairs $(R,E_{\alpha})$.
\end{definition}
\subsection{Neyman-Pearson Channel Discrimination Exponent}
Here we consider the channel discrimination problem for a given sequence of codebooks.
To this end, we define the conditional information divergence (or relative entropy) as 
\begin{equation}
D(W\|V|P_X) \triangleq  \sum_{x\in \mathcal{X}} P_{X}(x)D(W(\cdot|x)\|V(\cdot|x)) 
\end{equation}
where $D(W(\cdot|x)\|V(\cdot|x))$ is the information divergence between $W(\cdot|x)$ and $V(\cdot|x)$.
We start with the following characterization of the discrimination exponent for a given sequence of codewords. 
\begin{lemma}
\label{lm:stein_lemma}
Given a sequence of codewords $(x^n)_{n \in \mathbb{N}}$, for any $\alpha \in (0,1)$ we have
\begin{equation}
\label{eq:stein_lemma_eq}
    \liminf_{n \to \infty} - \frac{1}{n}\log \beta_{\alpha,n}(x^n) = \liminf_{n \to \infty} D(W\|V|\mathsf{P}_{x^n}).
\end{equation}
\end{lemma}
\begin{IEEEproof}
The above result follows from the generalized form of Stein’s lemma in [13, Theorem 1.2], as shown next.
We start by restricting $\psi_n(\cdot, x^n)$ to be a deterministic test, which is equivalently characterized by a decision region $\Omega$. 
Under this restriction, we know from \cite[Theorem 1.2]{csiszarInformationTheoryCoding2011} that for every $\alpha, \delta \in (0,1)$, we have
\begin{IEEEeqnarray}{c}
    \abs{-\frac{1}{n}\log \beta_{\alpha,n}(x^n) - E_n} \leq \delta
\end{IEEEeqnarray}
for sufficiently large $n$ (which depends on $\alpha$ and $\delta$), where 
\begin{IEEEeqnarray}{c}
    E_n \triangleq \frac{1}{n} \sum_{ i=1}^{n} \sum_{y \in \mathcal{Y}} W(y|x_i) \log \frac{W(y|x_i)}{V(y|x_i)} = D(W\|V|\mathsf{P}_{x^n}).
\end{IEEEeqnarray}
As a result, we know that for any $\alpha, \delta \in (0,1)$, we have
\begin{IEEEeqnarray}{c}
    \limsup_{n \to \infty} \abs{-\frac{1}{n}\log \beta_{\alpha,n}(x^n) - D(W\|V|\mathsf{P}_{x^n})} \leq \delta.
\end{IEEEeqnarray}
From this result, we obtain
\begin{IEEEeqnarray}{cl}
& \abs{ \liminf_{n \to \infty} -\frac{1}{n}\log \beta_{\alpha,n}(x^n) - \liminf_{n \to \infty} D(W\|V|\mathsf{P}_{x^n})} \\
    & = \abs{ \lim_{n \to \infty} \inf_{m \geq n}  -\frac{1}{m}\log \beta_{\alpha,m}(x^m) - \lim_{n \to \infty} \inf_{m \geq n} D(W\|V|\mathsf{P}_{x^m})} \\
    &= \lim_{n \to \infty} \abs{ \inf_{m \geq n} -\frac{1}{m}\log \beta_{\alpha,m}(x^m) - \inf_{m \geq n} D(W\|V|\mathsf{P}_{x^m})} \\
    \label{eq:Stein_exponent_proof}
    & \leq \lim_{n \to \infty} \sup_{m \geq n} \abs{-\frac{1}{n}\log \beta_{\alpha,n}(x^n) - D(W\|V|\mathsf{P}_{x^n})} \\
    & \leq \delta
\end{IEEEeqnarray}
where the inequality in \eqref{eq:Stein_exponent_proof} follows from Lemma \ref{lm:inf_less_sup}, given in Appendix \ref{appendix:Continuity_of_C}.
By making $\delta$ sufficiently small, we conclude that \eqref{eq:stein_lemma_eq} holds under the restriction that tests are deterministic. 
Finally, it follows from \cite[Problem 1.3]{csiszarInformationTheoryCoding2011} that randomized tests cannot help increase the exponent in this case. This concludes the  proof. 
%
%
%
\end{IEEEproof}

\begin{theorem}
\label{theorem:NP_exponent_given_codebook}
Given a sequence of codebooks $(\mathcal{C}_n)_{n \in \mathbb{N}}$, for any type I error threshold $\alpha \in (0,1)$ we have
\begin{IEEEeqnarray}{C}
    E_{\alpha} = \liminf_{n \to \infty} \min_{x^n \in \mathcal{C}_n} D(W\|V | \mathsf{P}_{x^n}).
\end{IEEEeqnarray}
\end{theorem}
\begin{IEEEproof}
The above theorem follows from Lemma \ref{lm:stein_lemma}. In particular, let $x^n(1) \in \mathcal{C}^n$ be such that
\begin{IEEEeqnarray}{C}
    x^n(1) = \arg\max_{x^n \in \mathcal{C}_n} \beta_{\alpha,n}(x^n).
\end{IEEEeqnarray}
Then we have the following lower bound
\begin{align}
E_{\alpha} & \triangleq \liminf_{n \to \infty} \min_{x^n \in \mathcal{C}_n} \frac{1}{n} \log \frac{1}{\beta_{\alpha,n}(x^n)} \\
& = \liminf_{n \to \infty}  \frac{1}{n} \log \frac{1}{\beta_{\alpha,n}(x^n(1))} \\
& = \liminf_{n \to \infty}  D(W\|V | \mathsf{P}_{x^n(1)}) \\
& \geq \liminf_{n \to \infty}  \min_{x^n \in \mathcal{C}_n} D(W\|V | \mathsf{P}_{x^n}).
\end{align}
On the other hand, if we let $x^n(1) \in \mathcal{C}^n$ be such that
\begin{IEEEeqnarray}{C}
    x^n(1) = \arg\min_{x^n \in \mathcal{C}_n} D(W\|V | \mathsf{P}_{x^n})
\end{IEEEeqnarray}
the we obtain the upper bound
\begin{align}
E_{\alpha} & \triangleq \liminf_{n \to \infty} \min_{x^n \in \mathcal{C}_n} \frac{1}{n} \log \frac{1}{\beta_{\alpha,n}(x^n)} \\
& \leq \liminf_{n \to \infty}  \frac{1}{n} \log \frac{1}{\beta_{\alpha,n}(x^n(1))} \\
& = \liminf_{n \to \infty}  D(W\|V | \mathsf{P}_{x^n(1)}) \\
& = \liminf_{n \to \infty}  \min_{x^n \in \mathcal{C}_n} D(W\|V | \mathsf{P}_{x^n}).
\end{align}
This completes the proof.
\end{IEEEproof}
\begin{corollary}
Define $E_{\alpha}^{\star} \triangleq \max_{(R,E_{\alpha}) \in \mathcal{R}_{\alpha}} E_{\alpha}$. For any $\alpha \in (0,1)$, we have
\begin{equation}
 E_{\alpha}^{\star} = \max_{P_X: \E_{P_X}[b(X)] \leq B}  D(W\|V | P_X).   
\end{equation}
\end{corollary}
\subsection{Trade-off Between Rate and Exponent}
We now present the rate-exponent trade-off under the Neyman-Pearson channel discrimination criterion. 
\begin{theorem}
\label{thm:disc_stein}
\(\mathcal{R}_{\alpha}\) for any $\alpha \in (0,1)$ is given by the set of all non-negative pairs  \((R,E)\) such that
    \begin{IEEEeqnarray}{Cl}
        & R \leq I(P_X, P_{Z|X}) \\
        & E \leq D(W\|V|P_X) 
    \end{IEEEeqnarray}
for some input distribution \(P_X\) on \(\mathcal{X}\) that satisfies \(\E_{P_X}[b(X)] \leq B\).
\end{theorem}
The above result can be proved using the exponent characterization in Theorem \ref{theorem:NP_exponent_given_codebook} and by following steps similar to the ones in the proof of  Theorem \ref{thm:disc_cost}. The proof is hence omitted.
\begin{example}
For the setting in Example \ref{example:1}, the rate-exponent region $\mathcal{R}_{\alpha}$ is given by
\begin{IEEEeqnarray}{cl}
    & R \leq H(\rho * p) - H(p) \\
    & E \leq \rho d(q\|1- q)
\end{IEEEeqnarray}
for some $\rho \leq B$, where $d(q\| 1-q)$ denotes the binary information divergence between $\bern(q)$ and $\bern(1-q)$.
For the setting in Example \ref{example:2},  we have $E \leq d(p\|q)$, which is still independent of the cost constraint.
\end{example}
\section{Conclusion}
The problem considered in this paper can be extended and generalized in several directions. 
For instance, one can think of extending the results to channels with general alphabets. In this case, while we do not expect the results to change, alternative proof techniques will be required. For instance, the achievability and converse  in Section \ref{subsec:proof_theorem_trade_off} are based on strong typicality and the method of types, which cannot be used in the case of general alphabets. 
Another extension is to consider, in addition to the coding rate and discrimination exponent, the channel coding exponent (i.e. reliability function) and to study the trade-off between all three. It may also be of interest for practical purposes to derive refined bounds that hold in the finite blocklength regime. 
Finally, one may also consider extending the setup to incorporate feedback and variable-length transmission. 
\appendices
\section{Continuity of $C(W \|V | P_X)$ in $P_X$}
\label{appendix:Continuity_of_C}

Here we show that under the assumptions that $W(y|x)V(y|x) \neq 0$ and $\abs{\mathcal{X}}, \abs{\mathcal{Y}} < \infty$, the mapping $C(W\|V | P_X)$, defined in \eqref{eq:def_C}, is continuous in the input distribution $P_X$  over the entire probability simplex $\mathcal{P}(\mathcal{X})$. 
In order to show this, we first prove the following auxiliary lemma.
\begin{lemma}
\label{lm:inf_less_sup}
Let $f(s)$ and $g(s)$ be real-valued functions  defined on the same domain $\mathcal{S}$. Then
\begin{IEEEeqnarray}{c}
    \abs{\inf_{s \in \mathcal{S}}f(s) - \inf_{s \in \mathcal{S}}g(s)} \leq \sup_{s  \in \mathcal{S}} \abs{f(s) - g(s)}.
\end{IEEEeqnarray}
\end{lemma}
\begin{IEEEproof}
First, note that 
\begin{IEEEeqnarray}{C}
    \sup_{s  \in \mathcal{S}} f(s) + \sup_{s \in \mathcal{S}} g(s) \geq \sup_{s  \in \mathcal{S}} (f(s) + g(s)).
\end{IEEEeqnarray}
It follows that
\begin{align}
 \sup_{s  \in \mathcal{S}} \abs{f(s) - g(s)} + \sup_{s  \in \mathcal{S}} g(s) & \geq    
 \sup_{s  \in \mathcal{S}} (f(s) - g(s)) + \sup_{s  \in \mathcal{S}} g(s) \\
 & \geq \sup_{s  \in \mathcal{S}} f(s).
\end{align}
In a similar fashion, we also obtain 
\begin{align}
 \sup_{s  \in \mathcal{S}} \abs{f(s) - g(s)} + \sup_{s  \in \mathcal{S}} f(s) & \geq \sup_{s  \in \mathcal{S}} g(s).
\end{align}
Combining the two, we get 
\begin{IEEEeqnarray}{c}
    \abs{\sup_{s \in \mathcal{S}}f(s) - \sup_{s \in \mathcal{S}}g(s)} \leq \sup_{s  \in \mathcal{S}} \abs{f(s) - g(s)}.
\end{IEEEeqnarray}
The lemma is then obtained after replacing $f(s)$ with $-f(s)$ and $g(s)$ with  $-g(s)$.
\end{IEEEproof}
We now define $g_s(x) \triangleq \log \left( \sum_{y \in \mathcal{Y}} W(y|x)^{1-s}V(y|x)^{s} \right)$
and the following inner product 
\begin{IEEEeqnarray}{C}
    \inner{P_X}{g_s} = \sum_{x \in \mathcal{X}} P_X(x) \log \left( \sum_{y \in \mathcal{Y}} W(y|x)^{1-s}V(y|x)^{s} \right).
\end{IEEEeqnarray}
Consider a sequence of distributions $P_X^{(n)}$ that satisfies $P_X^{(n)} \to P_X$ as $n \to \infty$. From Lemma \ref{lm:inf_less_sup}, we have
\begin{IEEEeqnarray}{cl}
\abs{ \lim_{n \to \infty} C(W\| V |P_X^{(n)}) - C(W\| V |P_X) } \IEEEnonumber 
& = \abs{ \lim_{n \to \infty} \min_{s \in [0,1]}  \inner{P_X^{(n)}}{g_s} - \min_{s \in [0,1]}  \inner{P_X}{g_s} } \\
    & = \lim_{n \to \infty}  \abs{ \min_{s \in [0,1]}  \inner{P_X^{(n)}}{g_s} - \min_{s \in [0,1]}  \inner{P_X}{g_s} } \\
    & \leq \lim_{n \to \infty} \max_{s \in [0,1]} \abs{ \inner{ P_X^{(n)} - P_X }{g_s} }\\
    &=0.
\end{IEEEeqnarray}
which proves continuity of $C(W\| V |P_X)$ in $P_X$.
\bibliography{ref}
\bibliographystyle{IEEEtran}

\end{document}